\documentclass[review]{elsarticle}

\usepackage{hyperref}

\journal{Advances in Space Research}

\usepackage{siunitx} 
\usepackage{graphicx}
\usepackage{caption}
\usepackage{subcaption}
\usepackage{epsfig,rotating,portland,fancyheadings,amsmath}
\usepackage{txfonts}
\usepackage{relsize}

\biboptions{authoryear}

\newcommand{\me}{Mini-EUSO}

\newcommand{\mefovdegalt}{44 $\times$ 44 sq. deg.}

\newcommand{\meresolution}{$\sim$ 6 $\times$ 6 \si{\km\squared}}
\newcommand{\megtu}{\SI{2.5}{\mu s}}
\newcommand{\melensdiameter}{\SI{25}{\cm}}

\newcommand{\Cpp}{C\nolinebreak[4]\hspace{-.05em}\raisebox{.4ex}{\relsize{-3}{\textbf{++}}}}
\newcommand{\scamfovkm}{227.9 $\times$ 227.9 \si{\km\squared}} 
\newcommand{\scamresolution}{200 $\times$ 200 \si{\metre\squared}}

\begin{document}

\begin{frontmatter}

\title{Secondary cameras onboard the \me~experiment: Control Software and Calibration}

\author[riken]{S. Turriziani\corref{mycorrespondingauthor}}
\cortext[mycorrespondingauthor]{Corresponding author}
\ead{sara.turriziani@riken.jp}

\author[riken,lul]{J. Ekelund}
\author[riken]{K. Tsuno}
\author[riken,infnrm2]{M. Casolino}
\author[riken]{T. Ebisuzaki}

\address[riken]{Computational Astrophysics Laboratory, RIKEN, Hirosawa 2-1, Wako-shi, Saitama 351-01, Japan}
\address[lul]{Lule\aa~tekniska universitet, 971 87 Lule\aa, Sweden}
\address[infnrm2]{INFN Sezione Roma Tor Vergata, Via della Ricerca Scientifica 1, 00133 Rome, Italy}

\begin{abstract}
\me~is a space experiment selected to be installed inside the International Space Station. It has a compact telescope with a large field of view (\mefovdegalt) focusing light on an array of photo-multipliers tubes in order to observe UV emission coming from Earth's atmosphere. Observations will be complemented with data recorded by some ancillary detectors. In particular, the \me~Additional Data Acquisition System (ADS) is composed by two cameras, which will allow us to obtain data in the near infrared, and in the visible range. These will be used to monitor the observation conditions, and to acquire useful information on several scientific topics to be studied with the main instrument, such as the physics of atmosphere, meteors, and strange quark matter.  
Here we present the ADS control software developed to stream cameras together with the UV main instrument, in order to grab images in an automated and independent way, and we also describe the calibration activities performed on these two ancillary cameras before flight.
\end{abstract}


\end{frontmatter}

\section{The \me~experiment}

\me~(Extreme Universe Space Observatory) is a space experiment to be operated inside the International Space Station (ISS) with the goal to study phenomena emitting UV light in the Earth's atmosphere \citep{new13}. Selected and funded by the Italian and Russian space agencies, it is supposed to fly in 2019 and it will produce high resolution maps of the Earth in the $300-400\si{\nano\metre}$ range. In addition, the experiment is expected to characterize atmospheric events in the ${ms-\mu s}$ range, such as lightnings and Transient Luminous Events (TLEs). It will also observe meteors and search for nuclearites, and monitor space debris: the latter activity constitutes the first step for in-situ testing of the debris remediation system proposed by \cite{Ebisu2015}. Data will be registered with ${320}{\mu s}$ and ${40}{ms}$ resolution for TLEs and meteors respectively. A more detailed description of \me~science can be found in \cite{new13}. 

\begin{figure}
\centering
\begin{subfigure}{\textwidth}
  \centering
  \includegraphics[width=0.9\linewidth,angle=-0] {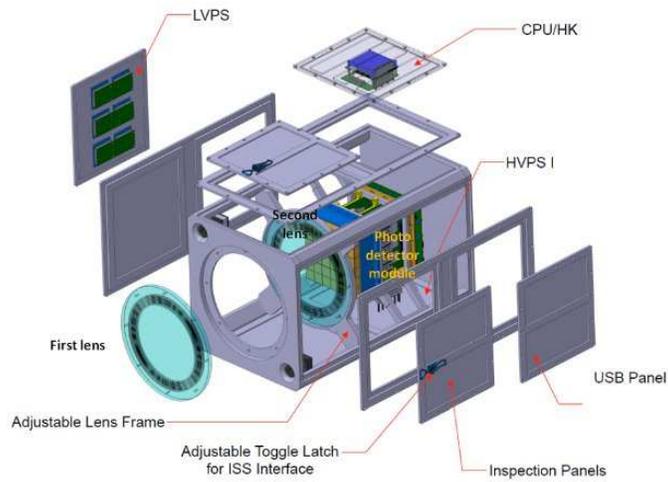}
  \caption{}
  \label{fig:sub1}
\end{subfigure}%

\begin{subfigure}{\textwidth}
  \centering
  \includegraphics[width=0.7\linewidth,angle=0] {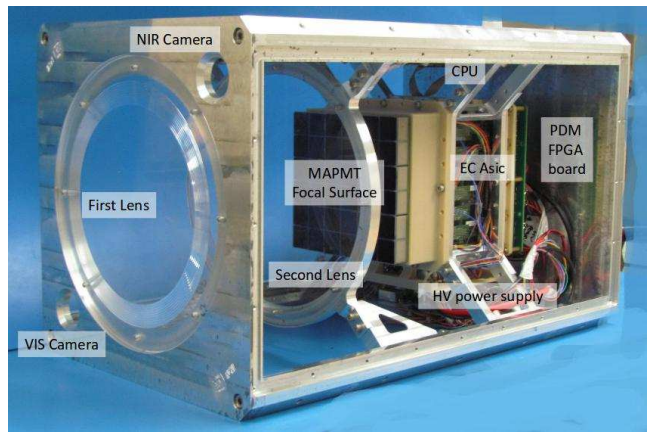}
  \caption{}
  \label{fig:sub2}
\end{subfigure}
\caption{(a): a 3-D model of \me , showing the main instrument and Fresnel optical system. Power and onboard data handling units are also located, namely the Low Voltage Power Supply (LVPS), the High Voltage Power Supply (HVPS), and the onboard processing unit (CPU/HK); (b): a picture of \me~with indicated the positions of the main subsystems. The SiPMs array is not shown as it is not a main subsystem.} 
\label{memodel}
\end{figure}

 The experiment has been built adopting a simple design (see Fig. \ref{memodel}): a small ($35\times 35 \times 60 \si{\cm\cubed}$) telescope focuses the incoming radiation on the Photo Detector Module (PDM), the main UV camera of \me. In particular, the telescope is a light weight ($\sim 30 \si{\kg}$) optical system which consists of two Fresnel lenses (diameter: \melensdiameter) and a UV filter (BG3), whereas the PDM is composed by 36 Multi-Anode Photo-Multiplier Tubes (MAPMTs), manufactured by Hamamatsu Photonics and arranged to form a square of $6\times6$ units or Elementary Cell (EC) units. Since each unit has 64 pixels, the full PDM has 2304 pixels. The units are powered by a Cockcroft-Walton High Voltage Power Supply \citep[HVPS;][]{hvps}. Given the optics and the altitude of the ISS, the PDM achieves a spatial resolution at ground of \meresolution. The signal from the PDM is then converted to digital by the Application Specific
Integrated Circuit (ASIC) boards, then processed and stored via onboard electronics subsystems. Another array composed by silicon photomultipliers (SiPMs) is located instead at the edges of the focal surface, in order to test this technology as a possible alternative for future missions within the JEM-EUSO framework \citep[JEM-EUSO, Joint Experiment Missions for Extreme Universe Space Observatory; see e.g.][]{jemeuso1, jemeuso2}. 

 Finally, two digital cameras, located at the edge of the front end of the telescope (outside the main optical system), constitute the so-called \me~Additional  Data  Acquisition  System (ADS). These two cameras, chosen to be sensitive in the infrared spectrum and in the visible range respectively, will acquire data independently of the PDM. These data will provide complementary information for both the study of transient phenomena and the measurement of Earth's emission. In particular, the combination of the observations from the PDM with NIR and VIS measurements will possibly provide a more complete characterization of meteors and bright fireballs, when they are detected by the different instruments. Moreover, we underline that to date no meteor has been observed in the wavelegth range covered by the NIR camera\citep{stme}. For a detailed description of the hardware composing the ADS, refer to Sect. \ref{cameras-hw}. 

 Given the absence of telemetry, observations will not be monitored in real time from ground, and data will be transferred to Earth, exchanging and trasporting the data disks back and forth from the ISS during resupply missions. \me~will be hosted inside the Russian module \textit{Zvezda} of the ISS, and placed to observe in nadir direction through the UV trasparent window. Grounding and power will be supplied by connection to the ISS. A low voltage power supply (LVPS), consisting of six DC/DC converters, produces the different voltage levels needed by the different subsystems. As the PDM will perform observations only during local night time on ISS, which corresponds to roughly 40\% of an orbit, a protective iris will be used. The astronauts' interactions with the experiment are supposed to be limited and devoted mainly to connecting the instrument, switching it on and off, and related to the disk exchange operations. This poses the requirement for a flight software capable to control all the instrument subsystems and data acquisition in a fully automated way. Furthermore, in order to optimize \me~scientific results, the software must also take into account the frequent day/night cycles of the ISS orbit to operate the detectors accordingly. 
  The signals of interest will be identified by a multi-level trigger algorithm, developed on purpose, which filters out noise  and implemented inside the dedicated Zync Field-Programmable Gate Array (FPGA) board \citep{new4}. The data fulfilling the requirements given by the triggers is sent to the onboard processing unit (labelled CPU or CPU/HK in the panels of Fig. 1), which saves the data on the disk. Moreover, the time resolution of the PDM (\megtu) permits to distinguish between fast and slow events, e.g. it can follow the shower development of Ultra High Energy Cosmic Rays (UHECRs) and study the emission of slower objects burning in the Earth's atmosphere (such as meteors, strange quark matter, and space debris). However, given the small diameter of the lenses, \me~can detect UHECRs only above the energy threshold of $E_{thr} \sim 1 \times 10^{21} eV$.

\section{Seconday Cameras: Hardware description}
\label{cameras-hw}

The specific hardware of \me~comprises both Commercial Off-the-Shelf components (COTS) and custom electronics boards developed by members of the JEM-EUSO collaboration. In particular, we adopted COTS products to assemble the ADS of \me, as this solution provided us several benefits such as lower initial costs, and less time spent on the development phase. 

For the ADS we opted to have two cameras, one sensitive in the visible band, and the other in the near infrared band, as the UV-transparent window in the Russian segment of the ISS does not permit to measure radiation emitted at longer wavelengths\footnote{This window has a fairly constant transmission function at a value of 86\% over a wide wavelength range, including the bands covered by the PDM, the VIS camera and the NIR camera. The transmission function drops abruptly to zero around 2500 nm, then it rises up at $\sim$ 50\% in a small band between $\sim$3000nm and $\sim$3400nm, dropping again to zero around 3500nm.}. The cameras are mounted outside the Fresnel lenses at the front of the main instrument (Fig \ref{memodel}). Both cameras are light (less than 40g each), compact and low power devices. They are connected to \me's embedded control system using USB 2.0 connectors, and they can be separately powered via their General Purpose I/O (GPIO) connectors. 

The visible camera (hereafter, VIS camera) was produced by FLIR Integrated Imaging Solutions Inc., formerly Point Grey Research.
 Its 1/3" sensor is based on CMOS technology and equipped with an additional infrared cut-off filter at wavelegths above 750 nm, making it sensitive in the $\sim$ \SI{400}{\nm}-\SI{750}{\nm} wavelength range. The VIS camera sensor is a color filter array designed using a Bayer filter mosaic, i.e. some pixels are sensitive to Red, some to Blue, some to Green light. In other words, each pixel in the acquired image is either Red, Green or Blue \citep[][; see Fig. \ref{bayer} for a visual representation]{manualVIS}. 

The near infrared camera (hereafter, NIR camera) was bought from Edmund Optics. The camera is actually a Black and White CCD Camera manufactured by FLIR, then a phosphorous coating was applied to its 1/3" sensor, and this coating converts near infrared photons to visibile range ones permitting the CCD to detect them. The coating phosphorous absorption response ranges from \SI{1500}{\nm} to \SI{1600}{\nm}, with two spectral peaks at \SI{1512}{\nm} and \SI{1540}{\nm}. Moreover, the NIR camera has an onboard
temperature sensor that can be used to monitor the ambient temperature within the camera case during operations \citep{manualNIR}.

Both cameras are equipped with a \SI{8.5}{\mm} Edmund Optics fixed focal length C-mount lens, yielding a $31.8^{\circ}$ Field of View for a 1/3" sensor, that corresponds to nearly \scamfovkm~on ground with a nominal pixel resolution of about \scamresolution~(when the camera is shooting at the maximum resolution). Each lens has a weight of 54g, a length of \SI{34.5}{\mm}, and no maximum working distance, extending performance out to infinity. Since the cameras support CS-mount lenses, CS to C Mount 5mm Spacer Adapters are added to obtain the correct focus.

Some more technical specifications of the cameras are given in Table \ref{cameras_table}. 

\begin{table}
  \begin{center}
  \caption{Mini-EUSO Cameras Main Characteristics}
  \label{cameras_table}
 {\scriptsize
  \begin{tabular}{|l|c|c|}\hline 
             & NIR Camera & VIS Camera \\ 
\hline
Camera Model & Chameleon 1.3 MP Mono USB 2.0 & Firefly MV 1.3 MP Color USB 2.0 \\ 
Sensor Name & Sony ICX445 & Sony IMX035 \\ 
Sensor type & CCD & CMOS \\ 
Sensor Size & 1/3" & 1/3" \\ 
Readout Method & Global Shutter & Rolling Shutter \\ 
Maximum Resolution & $1296 \times 964$ & $1328 \times 1048$\\
Standard Data Formats & $1280 \times 960$, $640 \times 480$ & $1280 \times 960$, $640 \times 480$\\
Maximum Frame Rate & 8-bit $1280 \times 960$ at 18 FPS & 8-bit (color) $1280 \times 960$ at 15 FPS\\
Pixel dimensions & $3.75$ \si{\micro\metre} $\times 3.75$ \si{\micro\metre} & $3.63$ \si{\micro\metre} $\times 3.63$ \si{\micro\metre} \\
Dimensions & 25.5 mm x 44 mm x 41 mm & 44 mm x 34 mm x 24.4 mm \\ 
Mass & 37 g & 37 g \\ 
Power Consuption & 2 W (max) at 5 V & $<$ 1 W\\

\hline
  \end{tabular}
  }
 \end{center}
\end{table}

\section{The ADS Control Software}

The ADS control software was developed in \Cpp~using the FlyCapture Software Development Kit (SDK). FlyCapture SDK gives a common software interface to manage and stream data from FLIR USB 3.1, GigE, FireWire, and USB 2.0 cameras using the same API under 32- or 64-bit Windows or Ubuntu Linux. We used Flycapture2 SDK (version 2.3.2.14) for 64-bit Linux \citep{manualAPI}. 
Control and status registers are programmed into each camera firmware, and these registers can be accessed to monitor or control each feature of the camera \citep{manualREG}. 

The ADS control software can be schematized as a sequence of three major blocks: connection, initialization, and streaming. We will describe each block in the following.

\subsection{Connection}

In the first major block the ADS control software checks how many cameras are connected to the system, and returns the number of detected cameras. The user can decide to stream both cameras or only one camera: a warning message will be issued in case the user asks to connect two cameras, but only one camera is connected, however the software will start with that one camera and stream data from it anyhow. Therefore, if this first stage completes successfully,  the software proceeds to the cameras initialization phase, whereas, if no camera is connected, the software exits with an error message.

\subsection{Initialization}

During this second stage, the ADS control software gets the working parameters to be used for each camera (i.e. \textsc{brightness}, \textsc{frame rate}, etc) from parameter files (hereafter, parfiles) and updates the values inside the register of each camera. 
 A list of possible parameters for FLIR cameras is given is Table \ref{par_table}.
We implemented a scheme with two parfiles for each camera: the software can choose between two different parameter files for initialization. They are: a) the \textit{default parfile}, which contains parameters validated and tested on ground before launch, and b) the \textit{current parfile}, to be used to observe with a different set of parameters with respect to our pre-launch fixed values. In fact, after the analysis on ground of the first sets of measurements, it could turn out that different parameters would better match the actual observing conditions on the ISS. In such a case, a \textit{current parfile} will be uploaded on the USB drive and used for the next observations. The software follows a hierarchical scheme: if the \textit{current parfile} is present, it is used for camera initilization, otherwise the software relies on the \textit{default parfile} only. 

However, right before getting through the initialization phase, some additional security checks are performed in order to validate the parfiles, and verify for example that the \textit{current parfile} is not trying to update a specific parameter with a value out of the allowed range. In case a problem is found, the camera register will be updated using the default values, and the software will prompt a warning to the user. 

Since each camera has its own parfiles, the value for parameters can be assigned autonomously for the two cameras. Then, according to what is parsed from the parfile, the control software will set to \textsc{on} or \textsc{off} each parameter, for example \textsc{frame rate}, \textsc{brightness} and so on; then, it will update the register with the value found in the parfile if the parameter is set to \textsc{on}.
Generally, the exposure time is defined by the combination of the values set for \textsc{frame rate} and \textsc{shutter}; however, in case the \textsc{frame rate} is set to \textsc{off} in the parfile, the \textsc{shutter} parameter only is used. 
Futhermore, the parameters \textsc{sharpness}, \textsc{saturation}, \textsc{trigger mode}, \textsc{trigger delay} \textsc{white balance}, \textsc{iris}, \textsc{hue},  \textsc{pan}, \textsc{tilt}, \textsc{gamma}, and \textsc{zoom} are always set to \textsc{off} in case they are supported by the camera. Note that not all the parameters are present in the camera register, as each camera can support different ones \citep[for more details, see][]{manualNIR, manualVIS, manualREG}.

\begin{table}
  \begin{center}
  \caption{Camera Parameters}
  \label{par_table}
 {\footnotesize
 
 \begin{tabular}{|l|l|}\hline
 Parameter & Description  \\ 
\hline

\textsc{autoexposure} & Used to automatically modify \textsc{shutter} or \textsc{gain} or both\\
\textsc{brightness} & Black level offset (\%)  \\
\textsc{frame rate} & Number of frames per second (fps)  \\
\textsc{gain} & Circuit gain of the sensor’s A/D converter used to transform the voltage of a pixel into a 12-bit value (dB)  \\ 
\textsc{gamma} & It determines the function between the level of incoming light and output picture \\
\textsc{hue} & Hue control (degree)  \\
\textsc{iris} & Lens aperture  \\
\textsc{pan} & Mechanism to horizontally move the current part of the sensor that is being exposed \\
\textsc{saturation} & It measures how distant a color is from a gray image of the same intensity (\%) \\
\textsc{sharpness} & Used for filtering the image in order to reduce blurred edges  \\
\textsc{shutter} & Integration time (ms)  \\ 
\textsc{tilt} & Mechanism to vertically move the current part of the sensor that is being exposed \\
\textsc{trigger delay} & Used to offset the synchronization of the camera to the trigger signal \\
\textsc{trigger mode} & Used to enable asynchronous trigger modes (hardware trigger, software trigger)  \\
\textsc{white balance} & Used to adjust color intensities to achieve more correct balance (only in color camera) \\
\textsc{zoom} & Zoom control \\

\hline
  \end{tabular}
  }
 \end{center}
\end{table}

\subsection{Streaming}

After successful connection and initialization, the software will start streaming the cameras at the same time, although not synchronized. Every single exposure is saved as a single image in 8-bit \textsc{raw} format at $1280 \times 960$ for both cameras. A \textsc{raw} format image contains only pixel values with no header or footer information. The software uses a \textsc{date} and \textsc{time} naming scheme for the images, getting the time of recording from \me's CPU clock. Moreover, a .log file is also recorded for each software run. This .log file records the two main output streams in Linux, i.e. standard output (\textsc{stdout}) and standard error (\textsc{stderr}). 

\subsection{Other characteristics}

The implementation of several security features makes the control software resistant to eventual hardware breakage/failure, i.e. signal handling (\textsc{sigterm}, \textsc{sigup}, \textsc{sigkill}) to exit gracefully, cleaning up the allocated resources, and the \textsc{bus reset}/\textsc{bus removal} catcher. In fact, if a camera gets eventually lost (due e.g. to a faulty cable), it will cause \textsc{bus reset} and \textsc{bus removal} events. The code stops and exits when a \textsc{bus reset}/\textsc{bus removal} event is detected, and the error trace also gives the serial number of the camera that caused the \textsc{bus removal}. This information is used then by the flight software to restart acquisition with the working camera only.  

Finally, there is a high level of flexibility on the path where to save images, as it has to be given by the user as input. This allows for storing images in external USB sticks or in the local disk of \me. 

The software (current version 4.3) is available on the GitHub platform: it is currently integrated into the flight software \citep{capel}, installed and successfully tested on \me~during laboratory tests, calibration and observation campaigns before flight.  

\section{Calibration of the cameras}

The first step in calibration was to estimate the bias and dark current for each camera.  The dark current is basically thermal noise causing some charge deposition in the pixels even in absence of illumination of the sensor, whereas the bias is the readout noise. which can vary from pixel to pixel. We found that VIS camera has a high bias (5 Analog to Digital Units, hereafter ADU) and a neglible dark current (less than 1 ADU) whereas NIR camera has a low bias (less than 1 ADU) but a high dark current, strongly dependent on the sensor temperature (less then 1 ADU at 310 K, rising to 2 ADU at 338 K). Note that we evaluated the dark current in ADU and ADU/s, and not in $e^{-}$ or $e^{-}/s$, because no good measurements of the ADC conversion gain is available for the sensors of the cameras in the experimental settings we operated.

In particular, for the dark current measurements we proceeded as follows. In order to avoid any stray light contamination, the cameras were put inside a small ($30 \times 8  \times 8$ cm) dark box.  We operated the cameras for more than an hour to allow them reaching the thermal equilibrium, in order to measure the dependence of the dark current as a function of the Shutter time at a roughly constant temperature. Both cameras were operated with gain set to 0 dB; moreover, since we set the \textsc{frame rate} to \textsc{off}, the \textsc{shutter} parameter coincides with the exposure time. Then, we acquired sets of 10 images each for 9 shutter times between 100 ms and 4000ms. The VIS camera has no internal sensor for temperature, however, since the two cameras were put together in the small box and operated at the same time, we reasonably assumed that the VIS camera has a temperature profile similar to the NIR camera, which does have a temperature sensor. The temperature measurements with the NIR camera showed a difference of about 1 K between the first and last set of images.
 
\begin{figure}
\centering
\begin{subfigure}{.5\textwidth}
  \centering
  \includegraphics[width=\linewidth,angle=-0] {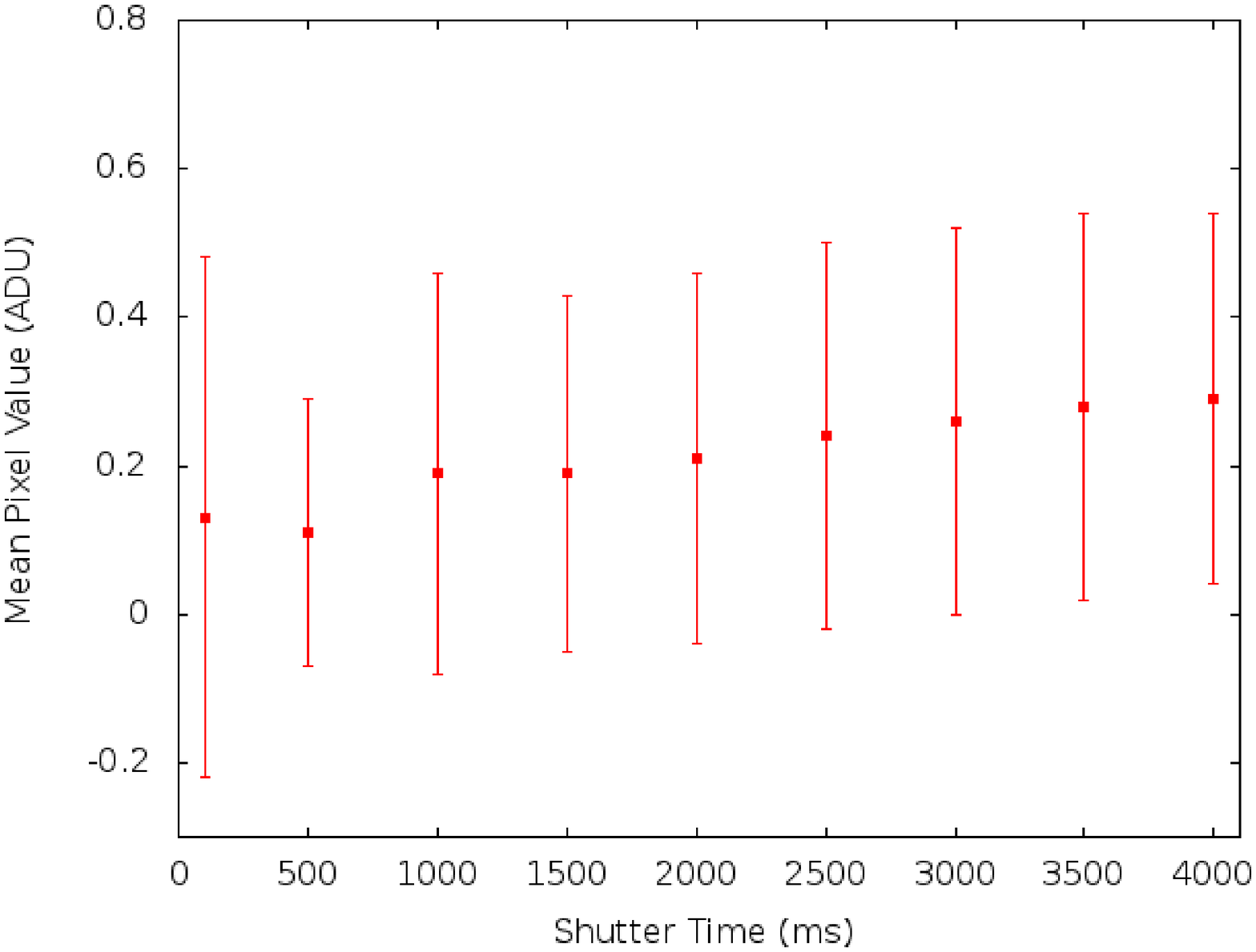}
  \caption{}
  \label{fig:subD1}
\end{subfigure}%
\begin{subfigure}{.5\textwidth}
  \centering
  \includegraphics[width=\linewidth,angle=0] {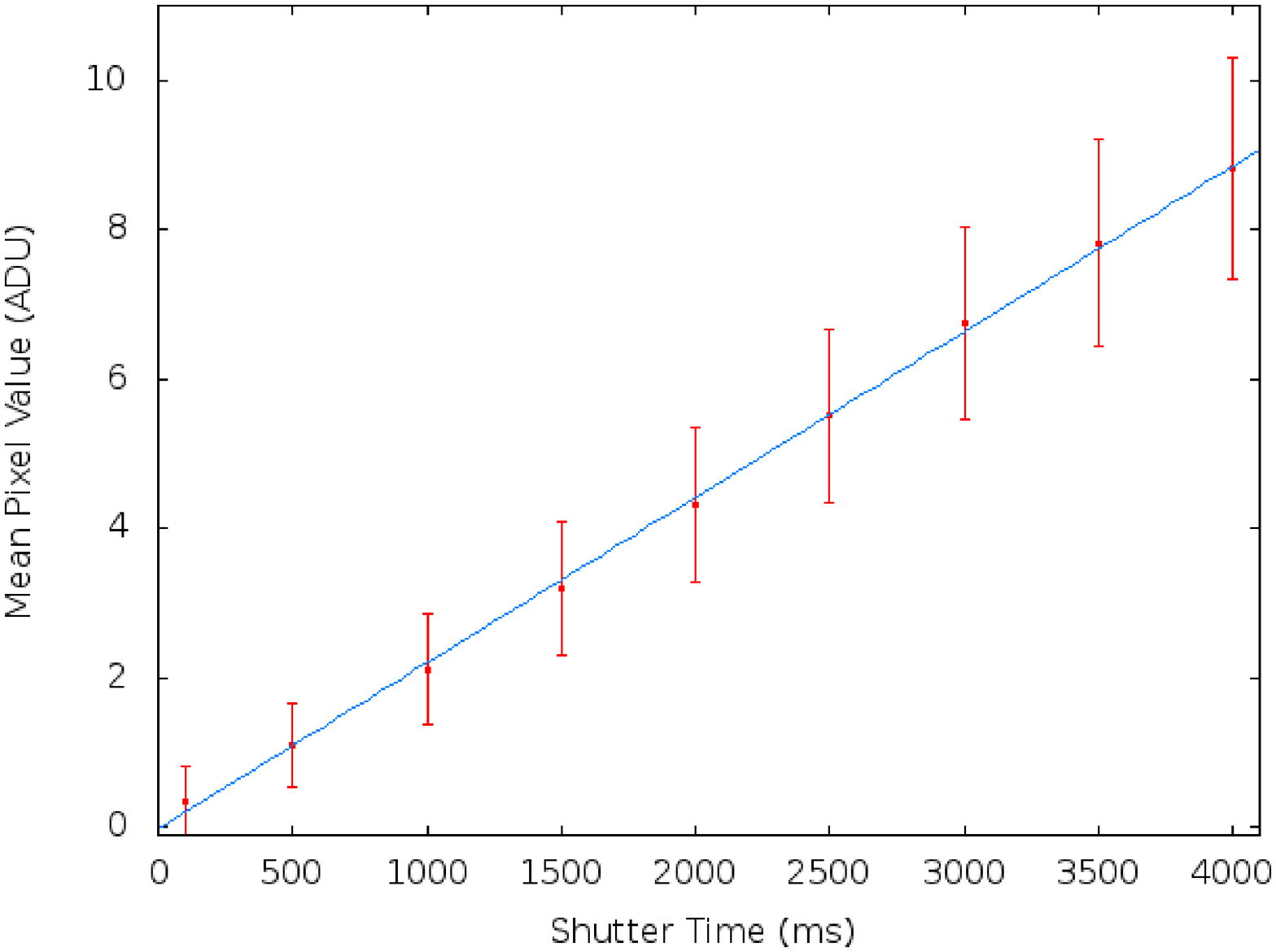}
  \caption{}
  \label{fig:subD2}
\end{subfigure}
\caption{(a): The dark current of the VIS camera versus Shutter time (ms); (b): The dark current of the NIR camera versus Shutter time (ms). } 
\label{dark1}
\end{figure}

Fig. \ref{dark1} reports the dependence of the dark current versus the Shutter time for each camera. The values in the plots are the pixel mean calculated over the entire image. As thermal noise is Poissonian, we estimated that the error for each pixel is the square root of the signal, so that the error bars in the plots of fig. \ref{dark1} are the mean, over the entire image, of the errors for each pixel. The left panel of Fig. \ref{dark1} shows that the dark current for the VIS camera is negligible. Then, we fitted the dark current for the NIR camera to a simple linear regression, $y=mx+ k$, giving $m=0.00222\pm0.00003$ and $k=-0.01\pm0.06$. We note that the $y$-axis intercept $k$ is consistent with zero from this fit, as shown in the right panel of Fig. \ref{dark1}. 

\begin{figure}
\centering
\begin{subfigure}{.5\textwidth}
  \centering
  \includegraphics[width=\linewidth,angle=-0] {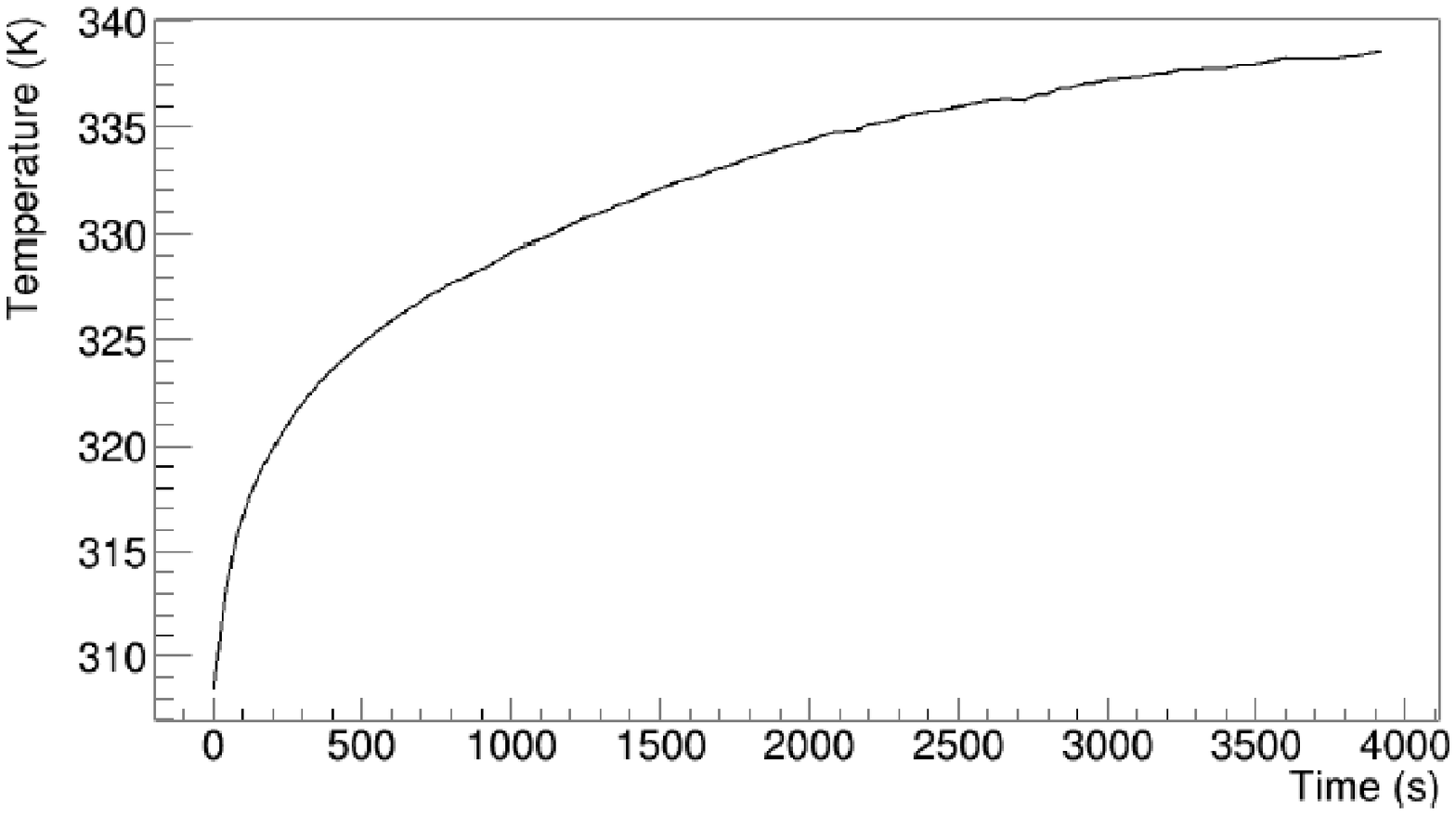}
  \caption{}
  \label{fig:subD3}
\end{subfigure}%
\begin{subfigure}{.5\textwidth}
  \centering
  \includegraphics[width=\linewidth,angle=0] {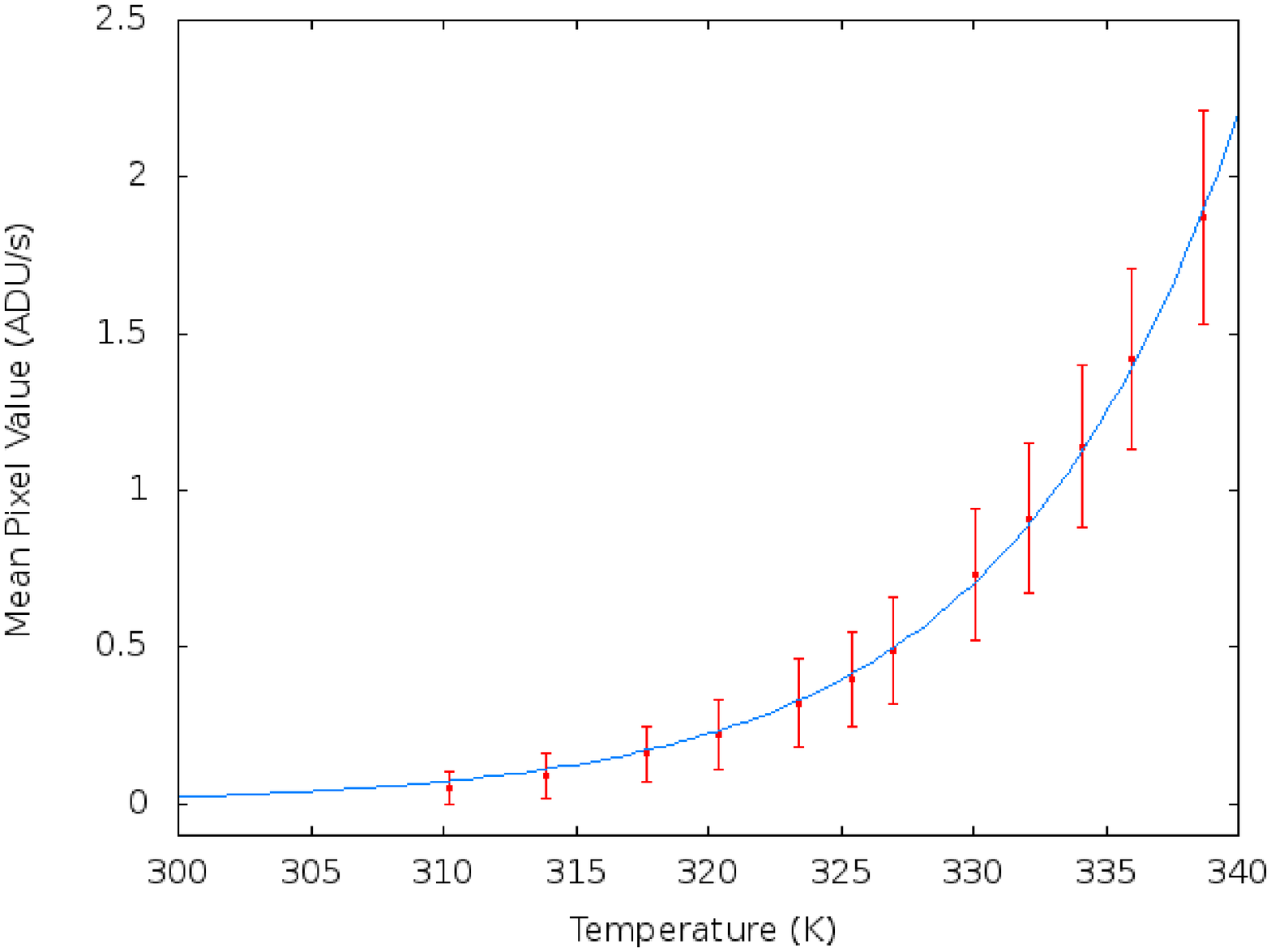}
  \caption{}
  \label{fig:subD4}
\end{subfigure}
\caption{(a): Temperature (K) of the NIR camera with respect to operation time; (b): The dark current of the NIR camera versus Temperature (K). } 
\label{dark2}
\end{figure}

A second laboratory test was performed with the cameras inside the small dark box to investigate the dependence of the dark current on the temperature: in this case, the cameras were operated to stream continuously and save images till they reached the temperature equilibrium.  Also during this test the cameras were operated with gain set to 0 dB. Assuming also in this case the same temperature profile for both cameras, we found that the VIS camera still has a very little amount of dark current in the images. Instead, we found a strong dependence of the dark current with temperature for the NIR camera.
We fitted an exponential function of the form $I_{dark} =  e^{a + b T}$ to the data, as shown in Fig. \ref{dark2}, where $I_{dark}$, $T$ is in K, $b$ in K$^{-1}$ and $a$ is an adimensional parameter. We found $a=-38\pm 1$ and $b=0.114\pm0.002$. Also in this case we calculated the mean dark current and relative error as described for the previous tests. 

To evaluate the bias frame, we acquired and averaged a number of images at the shortest possible exposure time time (0.1 ms for both cameras). We used the same experimental settings described above for the measurement of the dark current, i.e. the cameras were put inside the small dark box with their lenses covered.  

For more detailed calibration measurements, we relied instead on the Labspheres's LMS-760 Light Measurement Sphere\footnote{https://www.labsphere.com/site/assets/files/2794/pb-14022rev00\_lms.pdf}, an integrating sphere whose internal surface has been coated with a diffuse white reflectance material (Spectraflect\textsuperscript\textregistered). 
We used the LMS-760 located in the Optical Lab at National Institute of Polar Research (NiPR), Tachikawa, Japan, which
is a 2 m diameter integrating sphere. 
The sphere can be easily opened and closed since each hemisphere is mounted on a separate carriage. 
Inside the sphere there is a lamp mounting
bracket which can accept several different sockets for single contact and double contact lamps.
The system is controlled by a computer and a spectrometer measures the lamp emission. The maximum spectral 
brightness intensity is $30~ \mu W sr^{-1} m^{-2} nm^{-1}$ at 630 nm. 
The brightness is compatible to the order of 10 lux, i.e. visible objective illumination. We used a tungsten halogen lamp having
a temperature of 3100 K as its specification. 

We fixed the cameras on the mounting plate, we switched on the LMS-760 system, then we waited $\sim$ 30 minutes for it to warm up.
In the meantime, the cameras were connected to the testing laptop in order to make them become warm 
and reach the temperature plateaux. Before starting acquiring real data measurements with the control software, we used the 
FlyCap Demo software to decide the values for the parfiles via visual inspection of the images.
The lamp brightness was set at the maximum intensity during this preliminar evaluation stage.
Since the NIR camera proved to be less sensitive with respect to the VIS camera\footnote{Comparing the settings between the two cameras (NIR camera was operated with high gain and longer exposure at the maximum brightness with respect to the VIS during test 1 to get acceptable counts), we estimated that the sensor of the NIR camera is roughly 2.5 order of magnitudes less sensitive than the sensor of VIS camera}, we opted for a \textsc{gain} of 24 dB, even if this increases the dark current, whereas the \textsc{gain} was set to 0 dB for the VIS camera in all the performed tests. 

\begin{figure}
\centering
  \includegraphics[width=.8\linewidth] {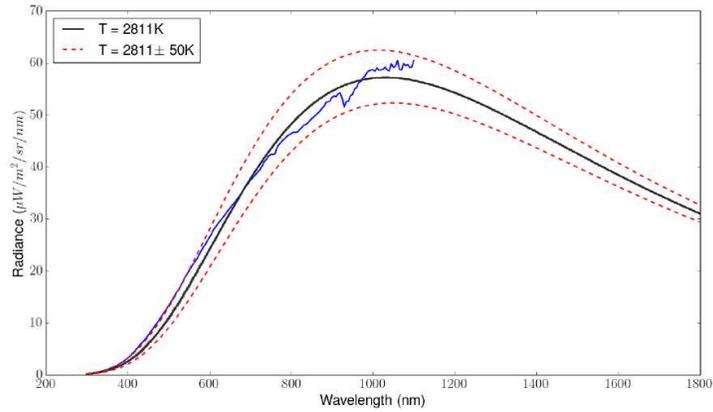}
  \caption{The measured spectrum of the lamp at the highest intensity used during calibration (blue line) compared with black body emission at different temperature.}
\label{bb}
\end{figure}

Fig. \ref{bb} shows a fit to the spectrum measured from the lamp. Note that the spectrum is measured for wavelengths
between $330$nm and $1100$nm. This covers the wavelength range of the VIS camera, but not that of the NIR camera. The spectrum was therefore fitted to Planck's law to extrapolate it for the NIR camera calibration purposes. 
From the fit we also inferred a slightly lower temperature for the blackbody (i.e. $\sim2811K$) with respect to the nominal value of 3100 K. 

For each measure we acquired the spectrum of the lamp and recorded its spectral brightness intensity. 
Each set of recorded frames is composed by 15 images.

\begin{itemize}

\item \underline{Test 1}: The cameras were acquiring data at the same time. The exposure time was set to 100 ms for the VIS camera, and 
to 4000 ms for the NIR camera. We acquired frames with cameras exposed at 7 different lamp illuminations, starting from the maximum brightness
available and decreasing its intensity. We acquired sets of dark frames for both cameras before and after each illumination measurement set.

\item \underline{Test 2}: This set of measurement was performed with the NIR camera only. We repeated Test 1, with the difference that this time we started from low brightness level and increasing it to the maximum brightness intensity. We performed this test only with the NIR camera because it is expected that the phosphor coating applied to the sensor could cause a non-linear behavior.  

\item \underline{Test 3}: The cameras were acquiring data at the same time. The lamp illumination was fixed at the maximum available 
brightness, and the cameras acquired sets of frames with 10 different exposure times, starting from 100 ms and increasing up to 4000 ms. 
We acquired sets of dark frames for both cameras before and after each measurement at a different exposure time.

\end{itemize}
 
\begin{figure}
  \includegraphics[width=\linewidth] {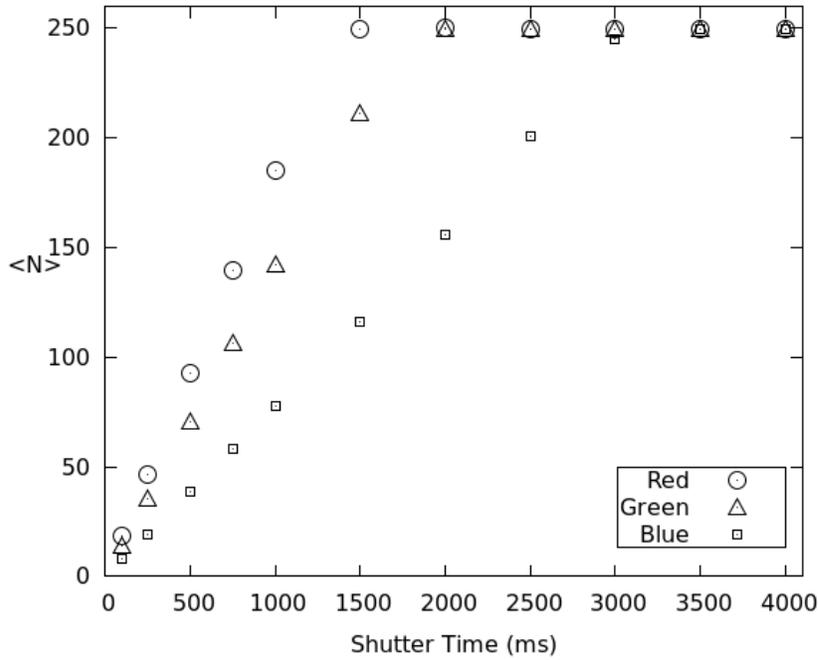}
  \caption{The mean pixel value $<N>$ is plotted versus shutter time (in ms) for the VIS camera. Data were acquired with lamp illumination fixed at the maximum available brightness ($\sim 10$ lux).}
\label{colors}
\end{figure}

These tests allowed us to determine that the VIS camera has good sensitivity for all the colors: Red, Green and Blue. 
We examined the saturation intensity for each color versus the Shutter times at constant illumination: we separated the pixels of different colors and we calculated the mean for each color. Fig. \ref{colors} shows the results for the maximum available brightness ($\sim 10$ lux) of the lamp: from this plot it would be easy to infer that the color which saturates faster is Red. However, the values shown in Fig. \ref{colors} do not take into account the spectrum of the lamp and the different sensitivity of the VIS camera for each color. Therefore, to study which color is the most sensitive to the incoming light, we integrated the intensities from the lamp spectrum taking also into account the sensitivity of the camera for each color. For the maximum brightness, we found that the integrated intensities are:  $\sim 1270\mu W~m^{−2}~sr^{−1}$, $\sim 1070\mu W~m^{−2}~sr^{−1}$, $\sim 650\mu W~m^{−2}~sr^{−1}$ for Red, Blue and Green pixels respectively. Therefore, the amount of light received by Red and Blue pixels was almost of the same order of magnitude, whereas Green pixels received only half the intensity of the Red pixels. This result implies that if the intensity of the incoming light is identical for the different colors, the Green pixels are expected to saturate before the other colors. In particular, we estimated that a Green pixel will register a value of 279 every second if it is exposed to an intensity of $\sim 1270 \mu W~m^{−2}~sr^{−1}$. This is higher than the maximum pixel value, which is 255, meaning that Green pixels would saturate when these observing conditions are met. Therefore, we concluded that the VIS camera is most sensitive to Green.

We used the measurements to estimate the time to saturation depending on intensity for Red, Green and Blue pixels, to have a reference about the observation conditions in which the pixels can be expected to be saturated, and in what order. During flight operation, we need to avoid to reach saturation which will cause loss of information about the phenomena we are interested to measure. Fig. \ref{VISsaturation} shows that the lowest possible intensity that can saturate any of the colors is $\sim 300\mu W~m^{−2}~sr^{−1}$. This intensity would saturate the Green pixels at a shutter time of 4000ms, i.e. the longest shutter time we can set for the VIS camera. The pixels of the two other colors saturate at slightly higher intensities: for the same shutter time, we have saturation at  $450\mu W~m^{−2}~sr^{−1}$ for the Red pixels, and $\sim 900\mu W~m^{−2}~sr^{−1}$ for the Blue pixels.
From this plot we can also infer the maximum intensity that the pixels can see before they saturate at the shortest shutter time ($\sim 0.1ms$: $\sim 1 \times 10^{7}\mu W~m^{−2}~sr^{−1}$ for Green pixels, $\sim 2 \times 10^{7}\mu W~m^{−2}~sr^{−1}$ for Red pixels and just below $\sim 4 \times 10^{7}\mu W~m^{−2}~sr^{−1}$ for Blue pixels.

\begin{figure}
  \includegraphics[width=\linewidth] {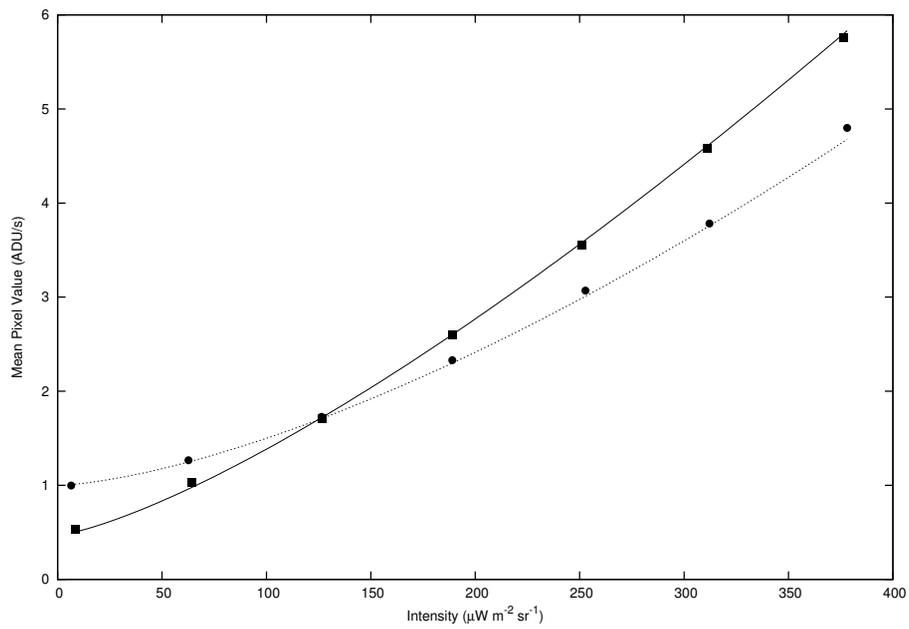}
  \caption{The plot show the fits to a power law of the mean pixel values of the images acquired during Test 1 (circles), and Test 2 (squares). 
 During Test 1 we acquired images while we were decreasing the intensity of the lamp, whereas during Test 2 we were increasing the intensity of the lamp. We integrated the intensities using the QE provided us by the vendor. See Table \ref{tab-fit} for parameters values.} %the text for more details, and
\label{nonlin}
\end{figure}

On the contrary, the sensitivity of the NIR camera is quite low: in fact, as stated before, we needed to set a high value for the ADC amplification gain. Nonetheless, the tests allowed us to assess the non-linearity of the NIR camera, likely due to the phosphor coating used to convert to visible light. To quantify the degree of non-linearity, we used data acquired during Test 1 and 2. We calculated the average of each set of 10 images acquired for each different intensity of the lamp. Then, we integrated the measured blackbody spectrum of the lamp to estimate the intensity over the wavelength range of the NIR camera. We used the quantum efficiency (QE) information provided us by the vendor to weight the integration at different wavelengths, even if this QE is probably affected by a large uncertainty. 
We fitted the data using a power law function of the form $y = a_{nir} x^{k_{nir}} + m_{nir}$, where y are the mean pixel values (ADU/s), x are the intensities ($\mu W  m^{-2}  sr^{-1}$), $a_{nir}$ is in $ADU~m^2  sr  μW^{-1}  s^{-1}$, $k_{nir}$ is an adimensional parameter and $m_{nir}$ is expressed in ADU/s. We show in Fig. \ref{nonlin} our fits for datasets acquired during Test 1 (circles) and Test 2 (squares). Table \ref{tab-fit} reports the results of the fits. The NIR camera shows a superlinear but subquadratic behaviour in both tests, as the index of the power law $k_{nir}$ is roughly 1.3 and 1.5 respectively. 

As stated prevoiusly in this Section, a non-linear behaviour is somehow expected, and depends most likely on the phosphor coating applied to the sensor in order to convert near-infrared photons to visible photons. \cite{LuLi} studied the non-linearity of phosphor coating for a CCD device at 1550nm, which is exactly in the wavelength range of our NIR camera, finding a good fit to sub-cubic power law (with index betwen 2.58 and 2.82). However, we cannot directly compare our results to \cite{LuLi} because the phosphor coating used for our NIR camera is proprietary, and the vendor did not disclose details on its manufacturing. 

\begin{table}
  \begin{center}
  \caption{Fit parameters for Fig. \ref{nonlin}}
  \label{tab-fit}
 {\footnotesize
  \begin{tabular}{|l|c|c|c|}\hline 
  Test & $a_{nir}$ & $k_{nir}$ & $m_{nir}$ \\            
       & ($ADU~m^2  sr  μW^{-1}  s^{-1}$) &   & ADU/s \\
\hline
Test 1 (circles in Fig. \ref{nonlin}) & $0.0005 \pm 0.0001 $  & $1.50 \pm 0.03$   & $1.00 \pm 0.03$\\ 
Test 2 (squares in Fig. \ref{nonlin}) & $0.0020 \pm 0.0001$ &  $1.33 \pm 0.01$  & $0.47 \pm 0.01$ \\ 
\hline
  \end{tabular}
  }
 \end{center}
\end{table}

\begin{figure}
  \includegraphics[width=\linewidth] {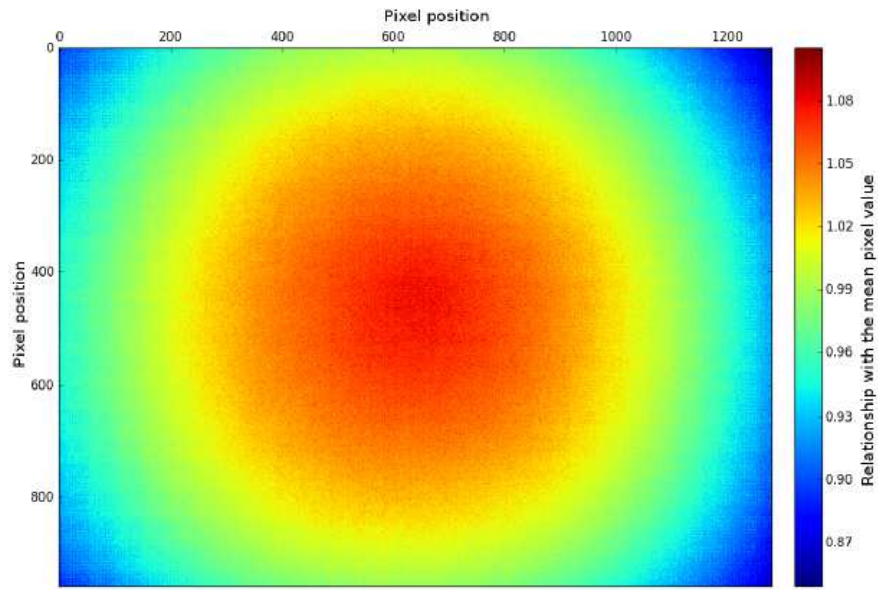}
  \caption{Normalized flat field for the VIS camera. The vignetting effect caused by the lens is clearly visible.}
\label{normffvis}
\end{figure}

\begin{figure}
  \includegraphics[width=\linewidth] {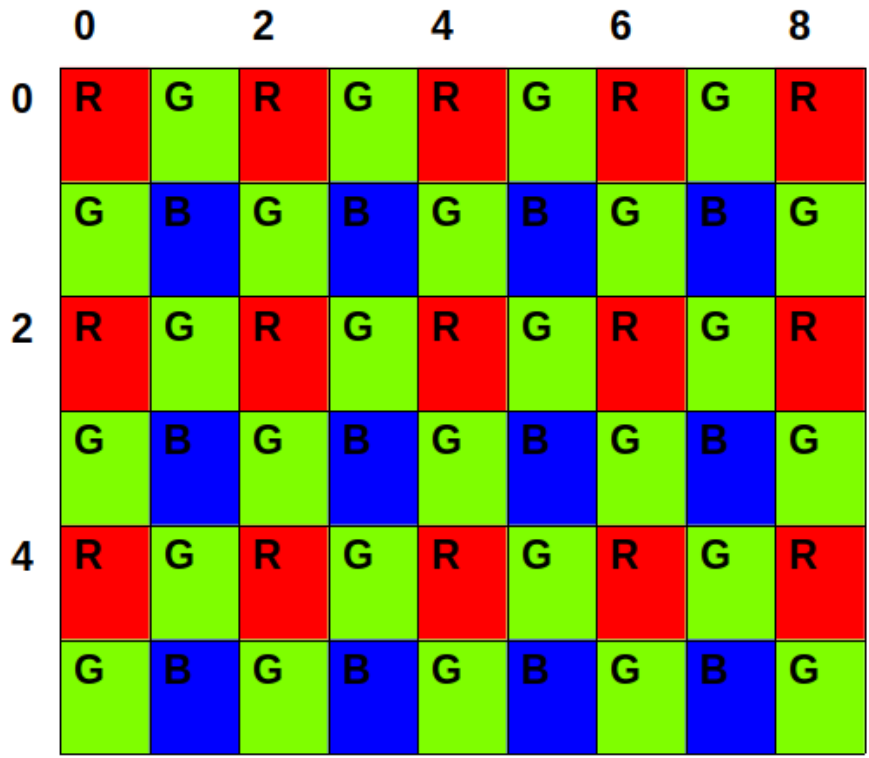}
  \caption{Representation of the Bayer pattern of the VIS camera. The x and y axes are the coordinates in the image with origin in the upper left corner. Each pixel is labelled as R for Red, B for Blue, and G for Green. The color of each square represents which color is registered by that specific pixel (colored Figure on the online version only).}
\label{bayer}
\end{figure}

\begin{figure}
  \includegraphics[width=\linewidth] {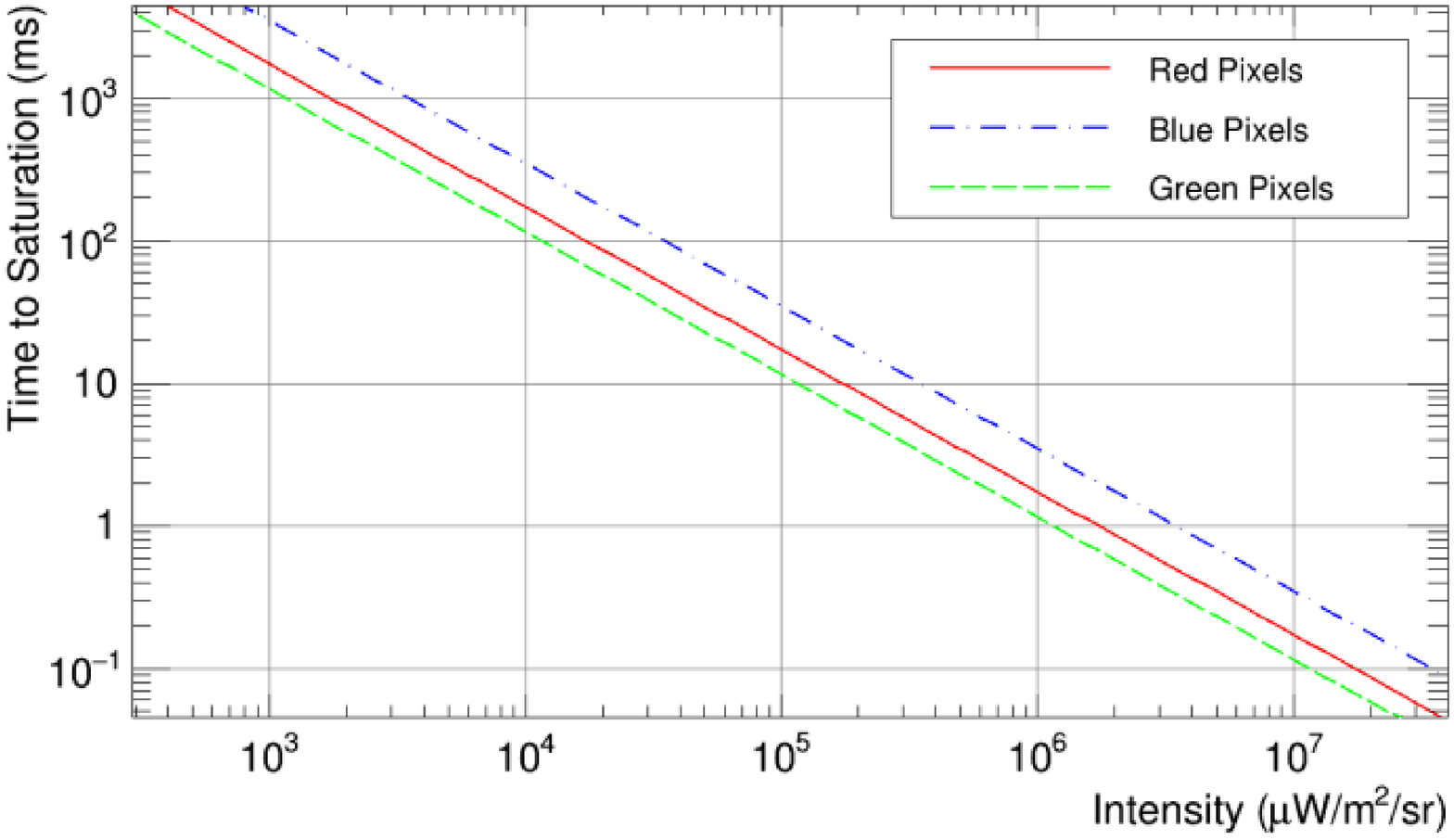}
  \caption{Time to saturation versus light intensity for the Red, Green and Blue pixels of the VIS camera. The highest value for the y-axis is about 4000ms as this is the longest possible shutter time.}
\label{VISsaturation}
\end{figure}

Finally, Fig. \ref{normffvis} shows the normalized flat field for the VIS camera, obtained using a flat field foil. We placed the camera with the aperture directly in contact with the surface in order to illuminate evenly the sensor. In Fig. \ref{normffvis} the vignetting effect caused by the lens is clearly visible. Unfortunately, a similar uniform light source was not available to be used in the case of the NIR camera, and even the data recorded with the light measurement sphere was not sufficient to determine a good flat field correction for it. 

To determine the value for parameters to be used inside the ISS, we applied the calibration results, taking into account two possible scenarios to operate the VIS and NIR cameras: 1. streaming only in conjunction with the observations of the PDM (i.e. night operation on the ISS, roughly 40\% of the orbit around Earth); 2. streaming continously (day and night), since the secondary cameras can perform observations also during daytime. We infer for the NIR camera that an exposure time of 4s would be the best choice to record possible signals of interest, especially for nighttime observations (when the illumination is expected to be lower), because we had to operate the camera with this long exposure time to get acceptable counts even at the lamp maximum brightness. 
On the contrary, to select the esposure time for the VIS camera, we should refer to Fig. \ref{VISsaturation} taking into account the emissions of events of interest and choosing an exposure time below the lines in order to avoid the saturation of the pixels, and thereby loss of information. However, we have to consider here an additional limit, posed by the available storage since each USB data disk has 512GB to save the data. We estimated that if we choose 1s for the VIS camera, and we are using 4s for the NIR camera, a data disk becomes full after two weeks for scenario 1, and after only one week for scenario 2. This implies that any shorter exposure time will fill the disk in much lower time, and will not be feasible for operations inside the ISS since the astronauts would have to exchange the disks much more frequently. From Fig. \ref{VISsaturation} we can see that, operating the VIS camera with 1s of exposure time, the Green pixels are expected saturate when the light illumination is $\sim 1 \times 10^{3}\mu W~m^{−2}~sr^{−1}$.

Therefore, the results from the calibration campaign, combined with the storage constraints for onboard data handling, suggest that the best exposure times to be used during \me~operations inside the ISS are 1s for the VIS camera, and 4s for the NIR camera.

\section{Conclusions}

\me~is a new experiment developed to measure from the ISS the Earth's UV light emission \citep{new13}, and it will be launched in 2019.   
We described here additional instrumentation that will contribute to complement the observations of the \me~main UV camera, namely the NIR camera and the VIS camera. A control software has been developed in \Cpp~to stream these secondary cameras together with the main UV camera of \me~so that they will acquire images in an automated and independent way. Finally, a calibration campaign has been performed on the additional cameras, and allowed us to define the best observational parameters for flight operations, which are 1s for the VIS camera, and 4s for the NIR camera.

\vspace{0.2cm}

\textbf{Acknowlegments}

ST was an International Research Fellow  of  the  Japan Society for the Promotion of Science.
ST would like to thank FLIR and previous Point Grey Technical Support Teams for useful discussions, 
especially Nobuya Okada, Brian Cha, Manuel Kroehl and Ana Florescu.  

The authors thanks M. Ricci and M. Bertaina for useful comments on the manuscript.

%-----------------------
This work was partially supported by the Italian Ministry of
Foreign Affairs and International Cooperation, Italian Space Agency (ASI)
contract 2016-1-U.0 "MINI-EUSO", State Space Corporation ROSCOSMOS, the Russian
Foundation for Basic Research, grant \#16-29-13065, and the Olle Engkvist
Byggm\"{a}stare Foundation.
%-----------------------

We dedicate this work to Dr. Yoshiya Kawasaki and Dr. Jacek Karczmarczyk, who passed away in 2016, 
and to Prof. Hiromichi Kamitsubo, who was the Godfather of the JEM-EUSO project and passed away in 2017.

%@arxiver{me-new.eps,dark-nir-new2.eps,Vsaturation.eps}
\end{document}